\documentclass[aps,twocolumn,showpacs]{revtex4}
\usepackage{graphicx}

\begin{document}
\title{Riemann-Silberstein representation of the Maxwell equations set}
\author{M. V. Cheremisin}
\affiliation{A.F.Ioffe Physical-Technical Institute,
St.Petersburg, Russia}
\date{\today}
\begin{abstract}
The complete set of Maxwell equations is represented by a single
equation using Riemann-Silberstein complex vector of electromagnetic field.
The consistent derivation of the Lorenz gauge condition is presented. We demonstrate that 
Fourier form of invariants ${\bf EB}$ and $E^{2}-B^{2}$ are
proportional to dissipated power and equal to zero respectively.

\end{abstract}
\maketitle

The relative orientation of
the magnetic and electric fields in a certain co-ordinate
system is of great importance since $E^2-B^2, {\bf EB}$ are
invariants upon Lorents\cite{Lorentz1904} relativistic transformation. In this
paper we represent the complete set of Maxwell equations by a single equation
using Riemann-Silberstein\cite{Silberstein1,Silberstein2} complex vector ${\bf F}={\bf
E}+i{\bf B}$ whose product contains the both invariants.

We start from the conventional Maxwell equations set:
\begin{eqnarray}
\text{rot}{\bf E}=-\frac{1}{c}\frac{\partial {\bf B}}{\partial t}
\qquad \qquad \qquad \text{div}{\bf B}=0,  \label{maxwell_1}\\
\text{rot}{\bf B} =\frac{1}{c}\frac{\partial {\bf E}}{\partial
t}+\frac{4\pi}{c}{\bf
J} \qquad \text{div}{\bf E}=4\pi \rho, \label{maxwell_2}
\end{eqnarray}
where the current ${\bf J}={\bf J}^{\text{free}} + c \cdot
\text{rot}{\bf M} + \frac {\partial \bf P}{\partial t}$ and the charge
density $\rho=\rho^{\text{free}}- \text{div}{\bf P}$ are generalized
with respect to those for free carriers ${\bf J}^{\text{free}},
\rho^{\text{free}}$ taking into account the polarization $\bf P$
and magnetization $\bf M$. In what follows we will search the
electromagnetic fields caused by preassigned charge and current densities. Thus,
we disregard the relationships ${\bf J}^{\text{free}}({\bf E}),{\bf P}({\bf E})$ and ${\bf M}({\bf B})$
known for electromagnetic fields in continuous media.

The Eqs.(\ref{maxwell_1}) leads to conventional definition ${\bf
B}=\text{rot}{\mathbf A}$, ${\bf E}= - \frac {1} {c} \frac
{\partial {\bf A}} {\partial t}-\nabla {\phi}$ known for vector ${\mathbf A}$ and scalar ${\phi}$ potential
respectively. With the help of the above notation Eqs.(\ref{maxwell_2}) yield
\begin{equation}
\fbox{} {\bf A} =- \frac{4\pi}{c} {\bf J}+\nabla {\psi}, \qquad
\fbox{} \phi =- 4\pi \rho- \frac{1}{c}\frac{\partial
\psi}{\partial t}.
\label{EM_potentials}\\
\end{equation}
where $\fbox{}$ is the d'Alembertian operator, then $\psi=\text{div}{\bf A}+\frac{1}{c}\frac{\partial
\phi}{\partial t}$ is a certain combination of potentials. Note that $\psi$ is arbitrary function
being, however, invariant upon Lorentz relativistic transformations similar to Maxwell's equations itself.
Nevertheless, in order to decouple Eqs.(\ref{EM_potentials}) an extra condition
\begin{equation}
\text{div}{\mathbf A}+\frac{1}{c}\frac{\partial
\phi}{\partial t}=0.
\label{Lorenz gauge}\\
\end{equation}
is usually implemented. The later is known as the gauge condition put forward first by Lorenz\cite{Lorenz1867}.
Historically, the latter was erroneously attributed\cite{Jackson2001} to H. A. Lorentz, who made, in turn,
the enormous contribution to classical electrodynamics. The crucial point of our paper concerns the rigorous
derivation of Eq.(\ref{Lorenz gauge}). We insist the later
arises from the Lorentz relativistic transformations of the potentials $\bf A, \phi$ counted for every
charge-centered coordinate system where the respective charge is assumed to be static.
\begin{figure}[tbp]
\begin{center}\leavevmode
\includegraphics[width=0.7\linewidth]{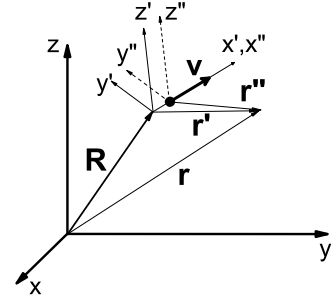} \caption[]{\label{Fig} Moving charge in laboratory L(bold), stationary L$'$(thin) and
charge-centered L$''$(dashed) coordinate frames.}
\end{center}
\end{figure}

Following \cite{Landau71} let us
consider(see Fig.\ref{Fig}) the elementary charge moving at a certain velocity ${\mathbf v}$ counted for laboratory coordinate
system L. In addition, we introduce a mobile coordinate system L$''$ moving synchronously with the charge.
The charge velocity is assumed to be directed along x$''$-axis of
the mobile frame. Then, let us make use of the stationary co-ordinate system
L$'$ whose axes coincide with those of the mobile frame L$''$ at a moment $t=0$.

We argue that the charge is motionless within the mobile frame L$''$. The respective electric field is described by
scalar potential ${\phi''}=\frac{e}{|r''|}$, where $\bf{r''}$ is the radius vector of the observation point
within the mobile frame L$''$. For stationary
coordinate system L$'$ both the scalar and vector potentials are present\cite{Landau71}
\begin{equation}
{\mathbf A'}=\phi'\frac{{\mathbf v}}{c},\qquad \phi'=\frac{\phi''}{\sqrt{1-v^{2}/c^{2}}}=\frac{e}{R^{*}}
\label{Charge potential}\\
\end{equation}
because of the Lorentz relativistic transformations\cite{Lorentz1904}. Here, $R^{*}=[(x'-vt)^{2}+(1-v^{2}/c^{2})(y'^{2}+z'^{2})]^{1/2}$,
where $x',y',z'$ denote the coordinates of the observation point counted for stationary frame L$'$, then $v=|{\mathbf v}|$.
We emphasize that the potentials specified by Eq.(\ref{Charge potential}) depend on the difference $x'-vt$. Neglecting
the spatial dependence of the charge velocity we obtain $\text{div}{\mathbf A'}=({\bf v},\nabla\phi')/c$ and, furthermore, recover
the Lorenz gauge condition $\psi'=0$ for sole charge within stationary frame L$'$. This result is valid overall the stationary
frame for arbitrary moment and charge velocity. Remarkably, the same condition remains valid for a sole charge within
laboratory frame L. Indeed, the stationary frame L$'$ can be superposed on the laboratory frame L under appropriate axes rotation and
subsequent translation(see radius vector $\mathbf R$ in Fig.\ref{Fig}) of L$'$-frame center. In general, the axis rotation and(or)
center translation of an arbitrary frame is known to keep the single Lame's indices\cite{Smirnov64}. The divergence of vector potential remains constant,
i.e. $\text{div} {\mathbf A}= \text{div} {\mathbf A'}$. Moreover, the scalar
potential remains unchanged as well, therefore $\phi=\phi'$. As a result, the Lorentz gauge is justified
for sole charge within the laboratory frame L as well.

Let us now evaluate the summation over the all elementary charges $\sum \limits_{j}\left[\psi_{j}=0\right]$, where the subscript denotes
$j$-th elementary charge. Obviously, for laboratory frame L we find both the vector ${\bf A}=\sum \limits_{j}{\bf A}_{j}$ and the scalar $\phi=\sum \limits_{j}\phi_{j}$ potentials being  find exactly those embedded into Eqs.(\ref{EM_potentials},\ref{Lorenz gauge}). With the help of
Lorentz gauge specified by Eq.(\ref{Lorenz gauge}), the Eqs.(\ref{EM_potentials}) obey a familiar form
\begin{equation}
\fbox{} {\bf A} =- \frac{4\pi}{c} {\bf J}, \qquad
\fbox{} \phi =- 4\pi \rho
\label{EM_potentials_simplified}\\
\end{equation}
and, moreover, result in conventional charge conservation law.

Let us discuss the gauge invariance conditions\cite{Landau71}.
In general, the electromagnetic fields ${\bf E},{\bf B}$ remain unchanged under the substitution ${\bf A}
\rightarrow {\bf A}+\nabla f$, $\phi \rightarrow {\bf \phi}-\frac{1}{c}\frac{\partial f}{\partial t}$, where $f$ is an arbitrary
scalar function. Searching the respective solution of Eqs.(\ref{EM_potentials_simplified}) one immediately find an extra
equation $\fbox{} f=0$ for presumably arbitrary function $f$. We attribute this finding to
potentials ${\mathbf A},\phi$ and electromagnetic fields${\mathbf E,B}$ are known to obey d'Alembert-like
homogeneous equation for charge-free media. We emphasize that the gauge invariance conditions
yield the replacement $\psi \rightarrow \psi +\fbox{} f$ and, hence provide the Lorenz gauge $\psi=0$ identity.

We now use the Riemann-Silberstein complex vector ${\bf F}={\bf
E}+i{\bf B}$ and, then make an attempt to re-write the Maxwell
equations set. At first, we note the identity
\begin{equation}
{\bf \dot{F}} ={\bf \dot{E}}-ic\cdot \text{rot} {\bf E},
\label{Riemann-Silberstein}\\
\end{equation}
which will be helpful in further investigation.

The Maxwell's equations set specified by Eqs.(\ref{maxwell_1},\ref{maxwell_2}) yields
\begin{equation}
\text{rot}{\bf F} =\frac{i}{c}({\bf \dot{F}}+ 4\pi {\bf J}),
\qquad \text{div}{\bf F}=4\pi \rho
\label{maxwell_3}\\
\end{equation}
or, in a shortened form
\begin{equation}
\fbox{} {\bf \dot{F}} = {\bf \ddot{I}}-ic \cdot \text{rot}{\bf \dot{I}},
\label{maxwell_4}\\
\end{equation}
where ${\bf \dot{I}}= 4\pi({\bf
\dot{J}}/c^2+ \nabla \rho)$ is an auxiliary vector. It is rather fascinating that the all Maxwell's equations may be
represented by means of a single equation. One can easily find the relationships
\begin{equation}
\fbox{} {\bf {E}}={\bf \dot{I}}, \qquad \fbox{} {\bf \dot{B}}=-ic\cdot \text{rot}{\dot {\bf I}}.
\label{maxwell_3}\\
\end{equation}
comparing Eqs.(\ref{Riemann-Silberstein}),(\ref{maxwell_4}).

We now able to discuss in greater details the problem of electromagnetic field invariants.
Within Fourier formalism ${\bf F}=\int \limits_{-\infty}^{+\infty}
{{\bf F}_{k,\omega}}e^{i\omega t-i{\bf kr}} \frac{d^{3}{\bf
k}}{(2\pi)^3} \frac{d\omega}{2\pi}$ the Eq.(\ref{Riemann-Silberstein}) yields the algebraic equation
\begin{equation}
{\bf F}_{k,\omega}={\bf E}_{k,\omega}+ \frac{ic}{\omega}[{\bf
kE}_{k,\omega}].
\label{RS-Fourier}\\
\end{equation}
which immediately gives the zero imaginary part of the product $\text{Im}{\bf
F}^2_{k,\omega}=2({\bf E}_{k,\omega},{\bf B}_{k,\omega}) \equiv 0$. In contrast, the real part of the product $\text{Re}{\bf F}_{k,\omega}^2={\bf E}^{2}_{k,\omega}-{\bf B}^{2}_{k,\omega}$ remains finite. Indeed, with the help of Fourier
form of Eq.(\ref{EM_potentials_simplified}),(\ref{maxwell_3}) we obtain
\begin{equation}
\text{Re}{\bf F}^2_{k,\omega}=\kappa^2({\bf
A}_{k,\omega}^{2}-{\phi}_{k,\omega}^{2}).
\label{field_invariants}
\end{equation}
where $\kappa^2=k^2-\omega^2/c^2$. According to Eq.(\ref{field_invariants}) the field invariant in terms of
Fourier components is proportional to certain combination complied from Fourier components of 4-vector potential.
This result is highly expected since the scalar ${\bf A}^2-\phi^{2}$ is known\cite{Stratton41}
to be invariant upon Lorentz relativistic transformations. Apart from Eq.(\ref{field_invariants}), a similar 
proportionality is occurred for Fourier replicas
of 4-vector current ${\bf J}^2-c^{2}\rho^{2}$ and scalar product $({\bf A},{\bf J})-c\phi\rho$ invariants\cite{Stratton41}.
We emphasize that Eq.(\ref{field_invariants}) can be, in addition, re-written in terms of dissipated power,
namely $\text{Re}{\bf F}^2_{k,\omega}=\frac{4\pi i}{\omega} ({\bf J}_{k,\omega},{\bf E}_{k,\omega})$. In absence 
of dissipation Eqs.(\ref{field_invariants}) gives a familiar result for electromagnetic wave
propagating in vacuum when ${\bf E}^{2}_{k,\omega}-{\bf B}^{2}_{k,\omega}\equiv 0$. The present study points to substantial
reduction of invariant multiplicity, at least within Fourier formalism. 

In conclusion, we represent set of Maxwell equations by single
equation using complex electromagnetic field. The Fourier form of
invariants $E^2-B^2$ and ${\bf EB}$ is proportional to dissipated
power and equal to zero respectively.

\end{document}